\documentclass[preprint,aps]{revtex4}
\usepackage{latexsym}
\usepackage{graphics}
\begin{document}
\title{On neutron number  dependence of B(E1;$0^+_1\rightarrow 1^-_1)$ reduced transition probability}
\author{R.V.Jolos$^{a}$, N.Yu.Shirikova$^{a}$, V.V.Voronov$^{a}$}
\affiliation{ $^{a}$Joint Institute for Nuclear Research, 141980
Dubna, Russia\\}
\date{\today}
\begin{abstract}
A neutron number dependence of the E1 $0^+_1\rightarrow 1^-_1$
reduced transition probability in spherical even--even nuclei is
analysed within the $Q$--phonon approach in the fermionic space to
describe the structure of collective states. Microscopic
calculations of the E1 $0^+_1\rightarrow 1^-_1$ transition matrix
elements are carried out for the Xe isotopes based on the RPA for
the ground state wave function. A satisfactory description of the
experimental data is obtained.
\end{abstract}
\maketitle

\section{Introduction}

The experimental data
\cite{Guhr,Zilges,Kneissl,Fransen,Robinson,Wilhelm1,Wilhelm2} demonstrate
that the lowest lying  1$^{-}_1$ states  in
spherical nuclei have mainly the structure of two--phonon states
arising as a result of the coupling of the collective quadrupole
$|2^+_1\rangle$ and octupole $|3^-_1\rangle$ states:
$|2^+_1\otimes 3^-_1 ;1^- M\rangle$. Thus, these states are isoscalar ones.
However, important information about their structure comes from the
E1 transitions which are characterized by the quantity
B(E1;$0^+_1\rightarrow 1^-_1)$. The operator of the E1 transition is mainly
an isovector operator and
this is a very important circumstance, as this means that analysing
E1 transition matrix elements we can obtain   information about
the proton--neutron structure of  $1^-_1$ excitations.

Among the important experimental facts characterizing strong E1 transitions
between the low--lying states  is the following one. There is a
minimum in the neutron number dependence of the matrix element
$|\langle 0^+_1\parallel {\cal M}\parallel 1^{-}_1\rangle |$
in the Nd, Sm
and Ba isotopes
when the number of
neutrons $N$ is equal to 78 or 86 \cite{Metzger1,Metzger2,Eckert}.
Such a behavior of
B(E1;$1^-_1\rightarrow 0^+_{1})$ as a function of the neutron number
was discovered earlier in the RPA--based calculations  \cite{Voronov}.
A very schematic IBA--based analysis of the E1 transition
$0^+_1\rightarrow 1^-_1$ \cite{Pietralla5,Smirnova,Jolos1}  showed that
the appearance of the minimum in the neutron number dependence of
B(E1;$0^+_1\rightarrow 1^-_1)$ is a result of  cancelation of  the proton
and neutron contributions to  an E1 transition matrix
element for a number of valence neutrons.
However, this phenomenological analysis cannot determine the number
of neutrons at which $|\langle 0^+_1\parallel {\cal M}\parallel 1^-_1\rangle |$
has a minimum. As it has been mentioned above, in Ba, Nd and Sm a minimum is located
in the isotopes with four valence neutron particles or holes. However,
the recently obtained results for the Xe isotopes \cite{Kneissl1} demonstrate
the absence of a minimum in the neutron number dependence of
$|\langle 0^+_1\parallel {\cal M}\parallel 1^-_1\rangle |$ when the number
of the valence neutron holes is equal to four. However, it is not improbable
that this minimum exists at larger numbers of the neutron holes where there are no data.

This is the aim of the present paper to analyse the problem of
appearance of a minimum in the neutron number dependence of
$|\langle 0^+_1\parallel {\cal M}\parallel 1^-_1\rangle |$
within the microscopic approach to description of the properties of the
$1^-_1$ state.

In our preceding paper \cite{Jolos1} we  analysed
the properties of the $1^-_1$ state in the framework of the fermionic Q--phonon
description of the low--lying positive and negative parity collective
states. The calculations performed  showed that the fermionic Q--phonon
approach was a good basis for the analysis of the properties of the
low--lying collective states of both parities. The consideration below is
based on this approach.

\section{Neutron number dependence of B(E1;$0^+_1\rightarrow 1^-_1)$ }

In the $Q$--phonon approach formulated for the fermionic configurational
space \cite{Jolos1} the $1^-_1$ state is presented by the
following expression
\begin{equation}
\label{5-eq}
|1^-_1, M\rangle ={\cal N}_{1^-_1}\left(\hat{Q}_{2}\hat{Q}_{3}\right)_{1M}|0^+_1\rangle ,
\end{equation}
where $|0^+_1\rangle$ is the ground state vector.
The expression for the normalization coefficient ${\cal N}_{1^-_1}$ is given in \cite{Jolos1},
$\hat{Q}_{2\mu}$ and $\hat{Q}_{3\mu}$ are  standard shell model quadrupole and octupole
moment operators.

Since mainly the spherical nuclei have been considered, it is assumed that
the ground states can be described in the RPA.
An approximation of the ground states wave vector by the RPA expression
should be discussed in more detail. Let us do it by the example of  Xe isotopes.
Heavier Xe isotopes are
spherical in their ground states. Therefore, for them an approximation
of the ground state by the RPA expression is well justified. The
lightest isotopes $^{124,126}$Xe are treated in IBM as belonging to the O(6)
dynamical symmetry limit. Let us compare qualitatively the structure of the
ground state wave vectors in the O(6) dynamical symmetry limit of IBM and
in the RPA of the microscopical nuclear model. In the O(6) limit of IBM
\begin{eqnarray}
\label{gsIBM}
|0^+_1\rangle =\sqrt{1-c^2_1 -c^2_2 - ...}\frac{1}{\sqrt{N!}}(s^+)^N|0\rangle\hspace*{2.0cm}\nonumber\\ +c_1\frac{1}{\sqrt{2}}(d^+d^+)_0\frac{1}{\sqrt{(N-2)!}} (s^+)^{N-2}|0\rangle\hspace*{1cm}\nonumber\\
+c_2 (d^+d^+d^+d^+)_0 \frac{1}{\sqrt{(N-4)!}}(s^+)^{N-4}|0\rangle +...,\hspace*{.5cm}
\end{eqnarray}
where $|0\rangle$ is the boson vacuum and $N$ is the maximum number of bosons.
In the RPA with accuracy sufficient for our discussion
\begin{eqnarray}
\label{gsRPA}
|0^+_1\rangle =\sqrt{1-c^2_1 -c^2_2 - ...}|0\rangle +c_1\frac{1}{\sqrt{2}}(A^+_2A^+_2)_0|0\rangle \nonumber\\
+c_2 (A^+_2A^+_2A^+_2A^+_2)_0 |0\rangle+...,\hspace*{1cm}
\end{eqnarray}
where $|0\rangle$ is the quasiparticle vacuum, $A^+_2$ is the operator
creating a collective superposition of two--quasiparticle states coupled to the angular momentum $L$=2.
We can say that there is a correspondence between
the bifermion operator $A^+_2$ of the RPA and the  boson operator
$d^+ s$ of the IBM. Therefore, the ground state wave functions in both approaches
have a similar structure. A difference can arise from the value of the
coefficient $c_1$ in (\ref{gsIBM}) and (\ref{gsRPA}). In the RPA the main
component of the ground state wave function is the first term in
(\ref{gsRPA}). However, in the IBM the second term in (\ref{gsIBM}) can give
a larger contribution. Using the consistent--Q IBM Hamiltonian we have
found that $^{128-136}$Xe can be described using the RPA ground state
wave vector. However, in the case of $^{124, 126}$Xe the RPA underestimates the
ground state correlations and a ground state can be described approximately as a mixture
of two lowest $0^+$ RPA states. As a consequence, the strength of the E1 transition from
the ground state will be fragmented between two $1^-$ states.

For the reduced matrix element of the E1 transition operator the following
expression was derived in \cite{Jolos1}
\begin{eqnarray}
\label{30-eq}
\langle 1^-_1\parallel {\cal M}(E1)\parallel 0^+_1\rangle =\left(B_p - B_n\right)e\cdot fm,
\end{eqnarray}
where $B_p$ and $B_n$ represent the proton and neutron contributions to the M1 transition matrix element, respectively.
The expressions for $B_p$ and $B_n$ are given in \cite{Jolos1}.

As it is seen from (\ref{30-eq}), the reduced matrix element
$\langle 1^-_1\parallel{\cal M}\parallel 0^+_1\rangle$ is equal to
the difference of the proton $B_p$ and neutron $B_n$ contributions. Let us consider
separately a neutron number dependence of $B_p$ and $B_n$. The results of
calculations for the Xe isotopes are shown in Fig.1. It is seen
that for the closed neutron shell (N=82) the neutron contribution to the E1
transition matrix element is sufficiently smaller than the proton
contribution. With increasing number of the neutron holes
the neutron contribution to the E1 transition matrix element increases
much faster than the proton contribution. However, the proton contribution
increases also due to an increase in the collectivity of the low--lying states.
As a result, a difference
$(B_p - B_n )$ decreases with the neutron number  and
$|B_p - B_n |$ takes a minimum value at $A$=128. With further decrease of
$A$ the E1 transition matrix element changes the sign and the modulus of the
difference $|B_p - B_n |$ increases again. Thus,
$|\langle 1^-_1\parallel{\cal M}\parallel 0^+_1\rangle |$ has a minimum at
$A$=128.
A similar picture can be observed in the Nd and Sm isotopes if we start
from the nucleus with the closed  neutron shell $N$=82 and then increase
the number of the neutron holes. Thus, in the semimagic nucleus with the
closed neutron shell the proton contribution to the E1 transition matrix
element is the largest one. With increasing the number of the valence neutrons
the neutron contribution increases. The proton and neutron contributions have
opposite signs  and for a number of neutrons
the absolute value of the E1 transition matrix element takes a minimum.
In the Nd and Sm isotopes this minimum takes place when the number of
the valence neutrons or neutron holes is equal to four. According to our calculations
in the Xe isotopes, this happens when the number of the neutron holes is
equal to eight. It is clear that this number can vary from element to element.
With further increase in the number of the neutron holes the absolute value
of the E1 transition matrix element in Xe isotopes increases again.
Note, however, that when
the neutron valence shell has a sufficiently large number
of the valence particles, i.e., when the neutron subshell is approximately
half filled, as in $^{108-116}$Cd$_{60-68}$ and
$^{116-124}$Sn$_{66-74}$ isotopes, the picture is changed.
The proton and neutron contributions vary more or less in parallel. This is
illustrated in Figs. 2 and 3.

The results of calculations of $|\langle 0^+_1\parallel {\cal M}\parallel 1^-_1\rangle |$
reduced matrix elements  for the Xe isotopes
are shown in Fig. 4 and in Table 1 together with
the experimental data from \cite{Kneissl1}. The calculated E1 transition
matrix elements decrease from $^{136}$Xe to $^{128}$Xe in agreement with
the experimental data. However, in lighter Xe isotopes the experimental
situation is unclear. Strong dipole transitions have been observed in these nuclei,
but the parity of the excited dipole states was not determined.
It is not improbable that they are due to the low--energy octupole strength
expected for these isotopes \cite{Zamfir}. If we assume that these
states are $1^-$, then we obtain from the experimental values of the
reduced ground state transition width $\Gamma^{red}_0$ the following
values of $|\langle 0^+_1\parallel {\cal M}\parallel 1^-_1\rangle |$:
0.051$e\cdot fm$ for $^{124}$Xe and 0.041$e\cdot fm$ for $^{126}$Xe.
The first value is close to the calculated one, the second one is somewhat higher.

As it is seen from Fig. 4 and Table 1, the calculated E1 transition matrix elements
in heavier Xe isotopes are
two times larger than the experimental ones. This can be explained in the following way.
The strength of the E1 transition $0^+_1\rightarrow 1^-_1$ correlates with a magnitude of the ground
state correlations. The stronger the ground state correlations the larger the E1 transition matrix element
$|\langle 0^+_1\parallel {\cal M}(E1)\parallel 1^-_1\rangle |$. The ground state correlations
increase with decreasing  energy of the low--lying collective states. For example, in the Xe
isotopes considered in this paper a number of the neutron holes in the valence shell increases
with decrease in the mass number $A$. Correspondingly, the energies of the $2^+_1$ and $3^-_1$ states
decrease and the ground state correlations produced by the quadrupole and octupole forces
increase with decreasing $A$. As a result, both the proton $B_p$ and neutron $B_n$ contributions
to $|\langle 0^+_1\parallel {\cal M}(E1)\parallel 1^-_1\rangle |$ increase with decreasing $A$.
It seems that in a semimagic $^{136}$Xe the ground state correlations in the neutron subsystem
are underestimated by the residual forces used. These residual forces overestimate an admixture
to the ground state wave function of the quasiparticle configurations with smaller energies and
underestimate  an admixture of the quasiparticle configurations with larger energies.
Only latter neutron quasiparticle configurations are presented in $^{136}$Xe because the neutron shell
is closed. As a result, the value of $B_n$ in $^{136}$Xe is underestimated and a difference
$(B_p - B_n )$ becomes too large.

In the Cd isotopes shown in Fig. 2 both the quadrupole and octupole ground state correlations increase with $A$.
As a result, $B_p$ and $B_n$ increase with $A$. In the Sn isotopes shown in Fig. 3 a proton contribution
$B_p$ is very small because a number of protons is semimagic. A neutron contribution $B_n$ is
almost independent of $A$. This happens because although the ground state correlations produced by
quadrupole forces increase with $A$ the correlations produced by octupole forces decrease with.

\section{Conclusion}

In this work a neutron number dependence of the $0^+_1\rightarrow1^-_1$
reduced transition probability in
spherical even--even nuclei is analysed on the basis of  the microscopic approach.
Our calculations for the Xe isotopes have demonstrated
that for the closed neutron shell (N=82) the neutron contribution to the E1
transition matrix element is relatively small in comparison with the proton
contribution. With increasing number of the neutron holes
the neutron contribution to the E1 transition matrix element increases
much faster than the proton contribution. As a result, a total transition
matrix element which is a difference
between the proton and the neutron contributions
decreases with the neutron number  and
takes a minimum value at $^{128}$Xe. With further decrease in
$A$ the E1 transition matrix element changes the sign and the modulus of
$\langle 1^-_1\parallel E1\parallel 0^+_1\rangle$
increases again. Thus, it follows from our calculations  that
$B(E1;0^+_1\rightarrow 1^-_1)$ has a minimum at $A$=128 for the Xe isotopes.
The number of the valence neutrons at which   $B(E1;0^+_1\rightarrow 1^-_1)$ has a minimum
can vary from element to element.
However, when the neutron subshell is approximately half filled
the picture is changed, i.e. the proton and neutron
contributions to E1 transitional matrix element vary approximately in
parallel without crossing.

The calculated E1 transition
matrix elements decrease from $^{134}$Xe to $^{128}$Xe in agreement with
the experimental data [2]. However, in lighter Xe isotopes the experimental
situation is unclear.

\section{Acknowledgment}

The authors are grateful to Prof. U.~Kneissl for illuminating discussions.
This work is supported in part by RFBR (Moscow), grant 04--02--17376.
RVJ thanks Humboldt Foundation for support.

\newpage


\begin{figure}[htbp]
 \resizebox{0.5\textwidth}{!}{%
  \includegraphics{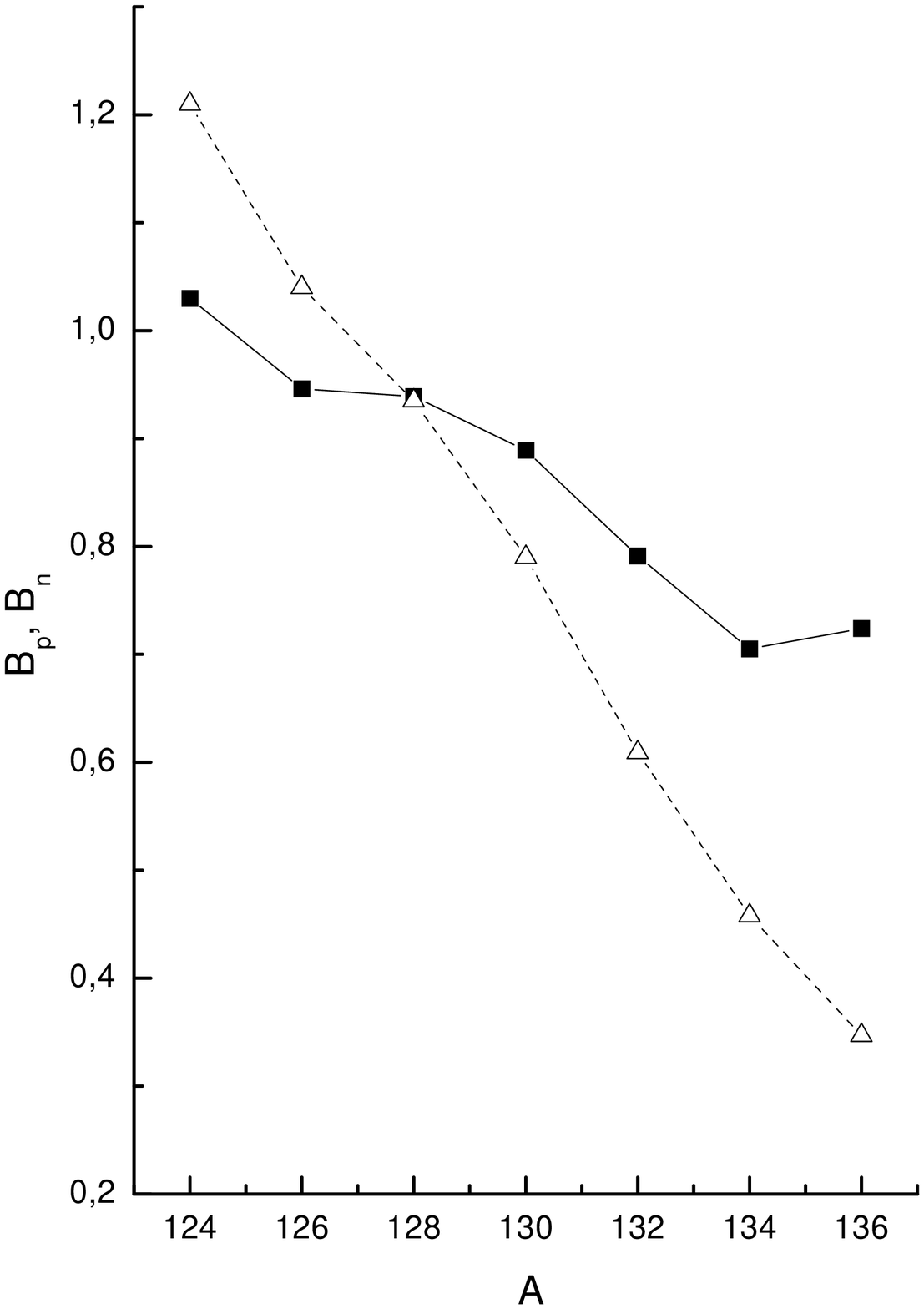}}
\caption{\label{fig:Xe}
Proton (solid line with full squares) and neutron (dashed line with open
triangles) contributions to the E1 transition  matrix element for the Xe
isotopes (in units 0.3 $e\cdot fm$).}
\end{figure}

\begin{figure}[htbp]
 \resizebox{0.5\textwidth}{!}{%
  \includegraphics{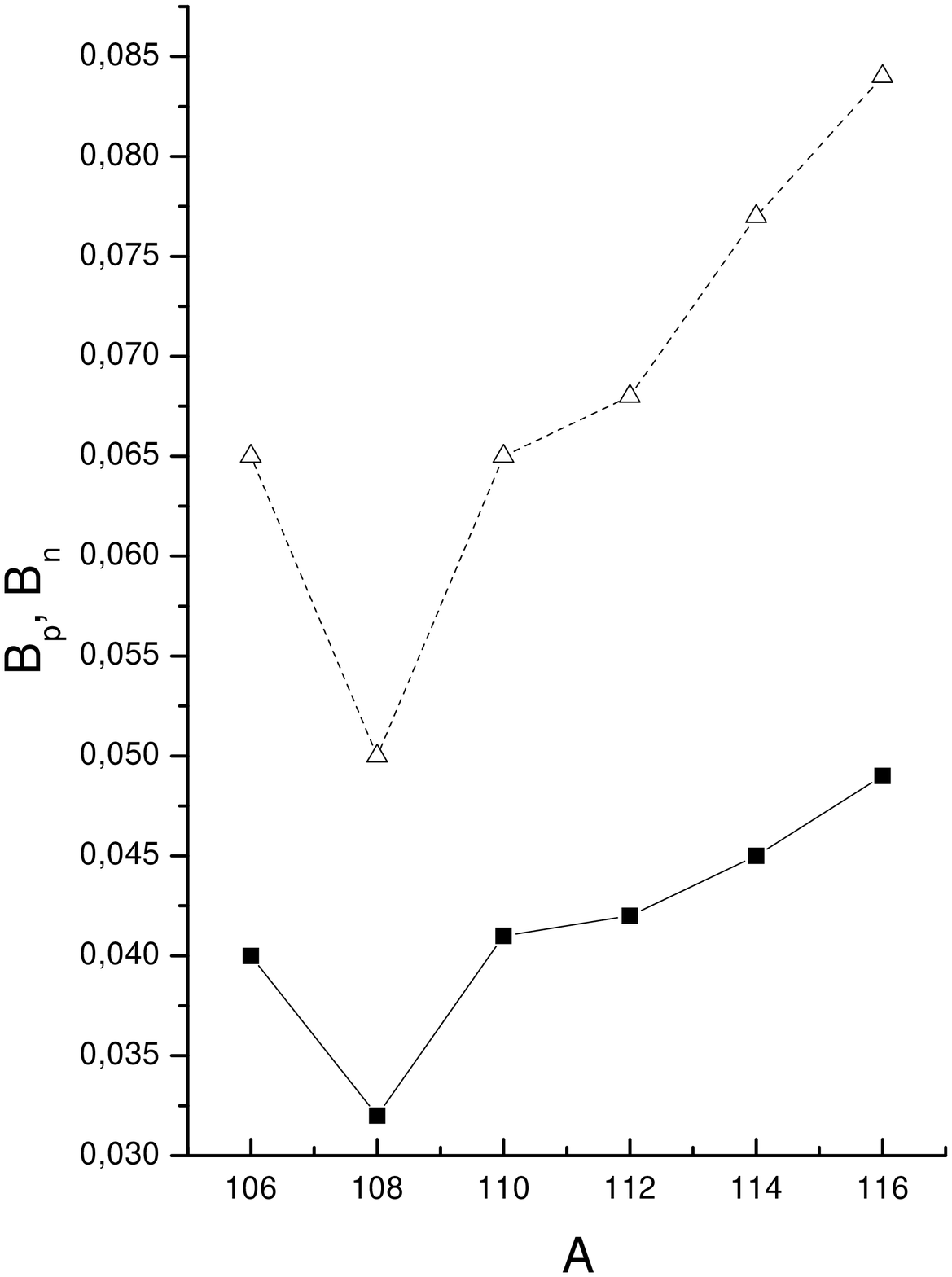}}
\caption{\label{fig:Cd}
The same as in Fig. 1 but for the Cd isotopes.}
\end{figure}

\begin{figure}[htbp]
\resizebox{0.5\textwidth}{!}{%
  \includegraphics{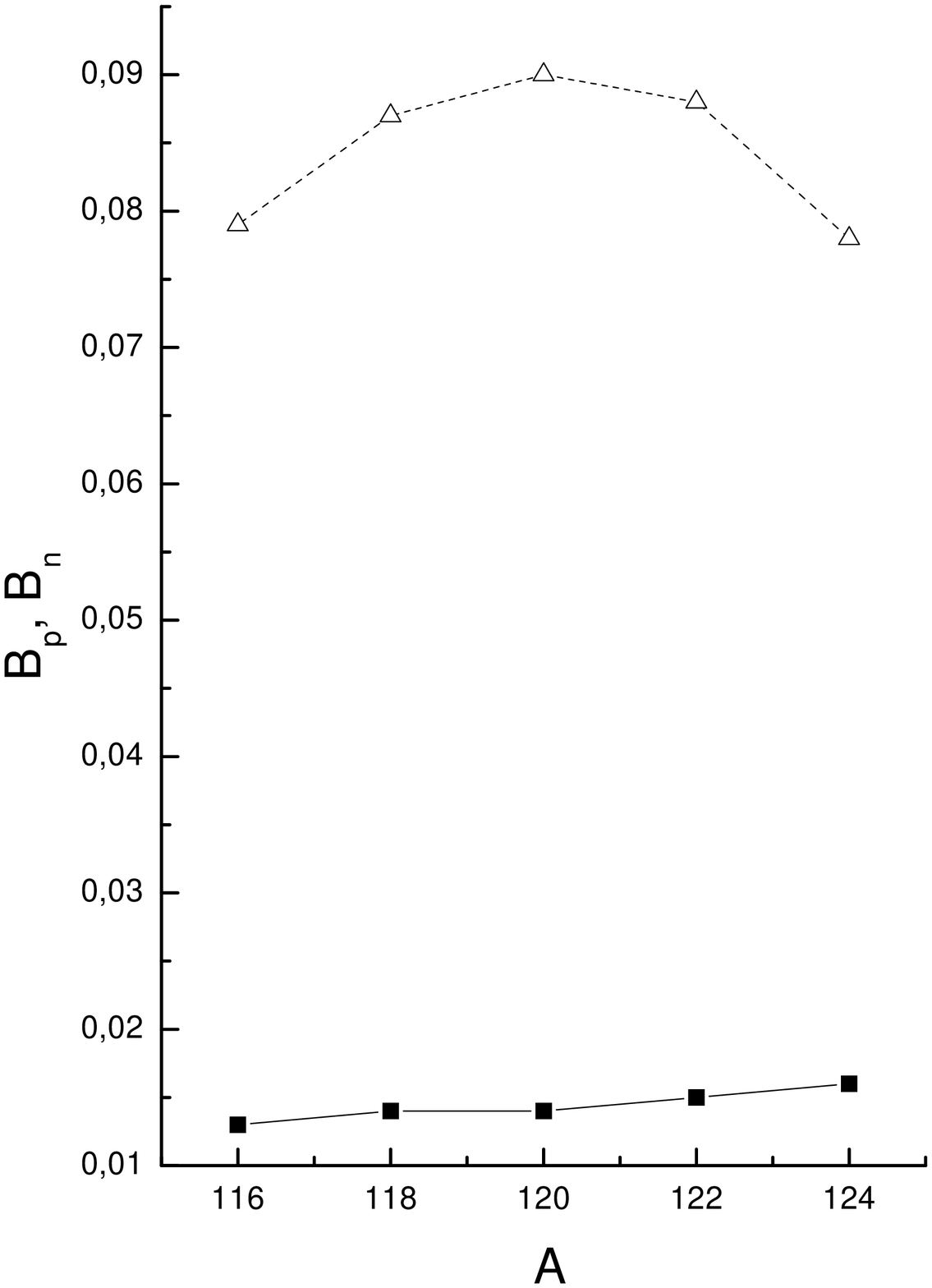}}
\caption{\label{fig:Sn}
The same as in Fig. 1 but for the Sn isotopes.}
\end{figure}

\begin{figure}[htbp]
\resizebox{0.5\textwidth}{!}{%
  \includegraphics{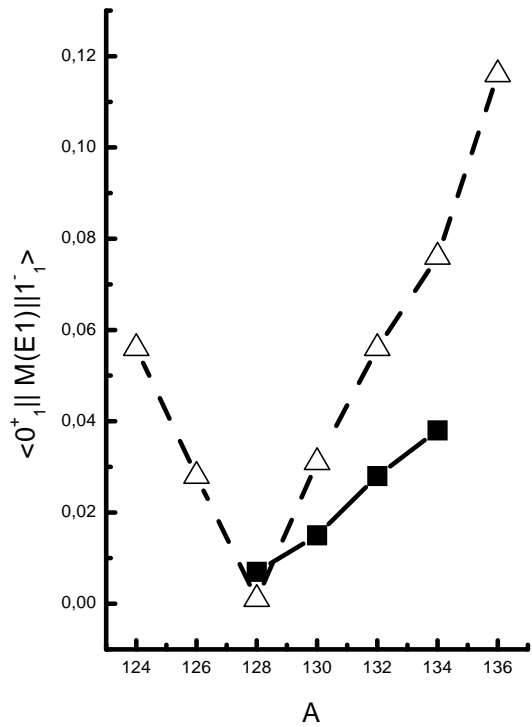}}
\caption{\label{fig:XeE1}
The experimental (solid line with open triangles) and calculated
(dashed line with full squares) electric dipole transition
matrix elements for the Xe isotopes.}
\end{figure}

\begin{table}
\caption{\label{tab:table1}
The experimental (exp) and calculated (calc) electric dipole transition matrix elements for the Xe
isotopes (in units $e\cdot fm$).
}
\begin{tabular}{ccc}
Nucleus  &\hspace*{0.2cm} $|\langle 1^-_1\parallel{\cal M}(E1)\parallel0^+_1\rangle |_{exp}$\hspace*{0.2cm} & $|\langle 1^-_1\parallel{\cal M}(E1)\parallel0^+_1\rangle |_{calc}$\hspace*{0.5cm}  \\
\hline
$^{124}$Xe & - & 0.056  \\
$^{126}$Xe & - & 0.028  \\
$^{128}$Xe & 0.007 & 0.001  \\
$^{130}$Xe & 0.015 & 0.031  \\
$^{132}$Xe & 0.028 & 0.056  \\
$^{134}$Xe & 0.038 & 0.076  \\
$^{136}$Xe & - & 0.116  \\
\end{tabular}
\end{table}

\end{document}